\documentclass[sigconf]{acmart}
\AtBeginDocument{%
  \providecommand\BibTeX{{%
    \normalfont B\kern-0.5em{\scshape i\kern-0.25em b}\kern-0.8em\TeX}}}

\copyrightyear{2025}
\acmYear{2025}
\setcopyright{cc}
\setcctype{by}
\acmConference[FAccT '25]{The 2025 ACM Conference on Fairness, Accountability, and Transparency}{June 23--26, 2025}{Athens, Greece}
\acmBooktitle{The 2025 ACM Conference on Fairness, Accountability, and Transparency (FAccT '25), June 23--26, 2025, Athens, Greece}\acmDOI{10.1145/3715275.3732071}
\acmISBN{979-8-4007-1482-5/2025/06}

\PassOptionsToPackage{table,xcdraw}{xcolor}
\usepackage{colortbl}

\newcommand{\edit}[1]{\textcolor{black}{#1}}

\begin{document}

\title[Secondary AI Stakeholders]{Secondary Stakeholders in AI: Fighting for, Brokering, and Navigating Agency}

\author{Leah Hope Ajmani}
\affiliation{%
  \institution{University of Minnesota}
  \city{Minneapolis}
  \state{Minnesota}
  \country{USA}}
\email{ajman004@umn.edu}
\author{Nuredin Ali Abdelkadir}
\affiliation{%
  \institution{University of Minnesota}
  \city{Minneapolis}
  \state{Minnesota}
  \country{USA}}
\email{ali00530@umn.edu}
\author{Stevie Chancellor}
\affiliation{%
  \institution{University of Minnesota}
  \city{Minneapolis}
  \state{Minnesota}
  \country{USA}}
\email{steviec@umn.edu}

\renewcommand{\shortauthors}{Ajmani, et al.}

\begin{abstract}
As AI technologies become more human-facing, there have been numerous calls to adapt participatory approaches to AI development—spurring the idea of participatory AI. However, these calls often focus only on primary stakeholders, such as end-users, and not secondary stakeholders. This paper seeks to translate the ideals of participatory AI to a broader population of secondary AI stakeholders through semi-structured interviews.  We theorize that meaningful participation involves three participatory ideals:  (1) informedness, (2) consent, and (3) agency. We also explore how secondary stakeholders realize these ideals by traversing a complicated problem space. Like walking up the rungs of a ladder, these ideals build on one another. We introduce three stakeholder archetypes: the reluctant data contributor, the unsupported activist, and the well-intentioned practitioner, who must navigate systemic barriers to achieving agentic AI relationships. We envision an AI future where secondary stakeholders are able to meaningfully participate with the AI systems they influence and are influenced by.
\end{abstract}

\ccsdesc[500]{Human-centered computing~Human computer interaction (HCI)}

\keywords{artificial intelligence, participatory ai, stakeholders, agency}



\maketitle
\section{Introduction}
\label{sec:introduction}
The recent proliferation of Artificial Intelligence (AI) has spurred calls to ``turn'' towards participatory approaches in AI development~\cite{Delgado2023-bo,Birhane2022-nd, Cooper2024-xg} called participatory AI (PAI). As its name suggests, PAI is the application of participatory design methods~\cite{muller1993participatory} to AI technology development, and is rooted in ethical, just, and community-oriented work~\cite{Kuo2024-nx, Bondi2021-lp, Tang2024-pb}.

A core question to PAI is about approach: \textit{what does meaningful AI participation look like?} This question is especially important to answer as new AI advancements transform our basic notions of ethics~\cite{Wu2023-ak}, justice~\cite{Ajmani2024-qk}, and human-centeredness~\cite{Chancellor2023-gd}. Prior PAI work relies on articulating current participatory practices~\cite{Delgado2023-bo} and their ability to moderate power imbalances~\cite{Corbett2023-yi}. In answering these questions, we can also understand the challenges and problems associated with current PAI -- such as superficial participant engagement (i.e., participation-washing~\cite{Sloane2022-va}). These works are crucial to understanding and evolving practices, especially as AI technologies are rapidly developing. 

However, participation is about more than just approach. Participatory AI also raises person-centric questions: \textit{who are we considering when engaging in participatory AI practices?}  In traditional participatory design, there is often an end-user group or affected community in mind~\cite{Suresh2022-nd, Stapleton2022-bu, Sanders2002-tn}. The subjects in these participatory approaches are often \textbf{primary stakeholders} -- those who are clearly and directly affected by the system being developed. For example, participatory AI work about child welfare algorithms often engages with social workers, affected parents, and government officials (i.e., end users, affected communities, and regulators)~\cite{Stapleton2022-bu, Brown2019-hc, Saxena2020-kd}. This work has proven overwhelmingly successful at improving specific algorithms for specific stakeholder groups.

As AI fundamentally transforms who is allowed to be a stakeholder, there are unique concerns over participating with secondary stakeholders. \textbf{Secondary stakeholders} are those who are influenced by and have an influence on AI systems, but they have indirect relationships with these systems. Formally defined in organizational sciences~\cite{Clarkson1995-we}, we articulate secondary stakeholders as those who (1) influence AI systems broadly but (2) lack a direct contract with an AI system. Previous work has theorized that, due to this indirectness, secondary stakeholders are difficult to reason about in the context of PAI~\cite{Sloane2022-va, Chancellor2023-gd, Li2023-wa}. However, if left unchecked, our current AI practices risk fueling the frustrations of secondary stakeholder groups.

Hollywood writers went on strike due to AI threats to their jobs~\cite{Wilkinson2023-mr}. Many people contribute to AI systems simply by being online~\cite{Li2023-wa}. Reddit moderators recently turned over $8,000$ subreddits private in protest of Reddit Inc.'s new data relationships with AI companies~\cite{Paul2023-lp}.  In all of these incidents, the main stakeholder group is secondary; they have a murky and indirect relationship with the AI system. Yet, their frustrations are real and urgent. Our work relies on the fundamental premise: \textbf{an ethical AI future is one where everyone gets their due data rights.} 

To achieve this future, we argue for PAI work to consider secondary stakeholders with the same ideals we apply to primary stakeholders. As AI systems grow in scale, the number of people who are secondary stakeholders grows accordingly. The term stakeholder now includes the many examples mentioned above -- from affected communities (in the context of welfare algorithms) to Hollywood writers, Reddit users, and moderators across the globe. Given the current frustrations that cut across these stakeholder groups and evidence that they are often exploited in AI contexts~\cite{Sloane2022-va, Li2023-wa}, secondary stakeholders warrant increased attention from PAI practitioners. We need to actively translate PAI ideals from primary stakeholders to secondary ones. The first step here is to understand current barriers to achieving this translational effort.

In this paper, we empirically investigate the unique challenges of being a secondary stakeholder in an AI system. We start from theoretical foundations previously established at FAccT: due data rights in AI contexts require agency over one's data~\cite{Ajmani2024-qk}. We expand this premise beyond data agency and describe meaningful AI participation as a variation of Arnstein's Ladder of Participation from civic engagement~\cite{Arnstein1969-bb}. Meaningful participation is the step-by-step process of creating (1) informed, (2) consensual, and (3) agentic relationships with AI systems. With the ladder as our guiding lens, we present the results from semi-structured interviews with 12 secondary AI stakeholders about their experiences and frustrations with AI. We combine our interview findings into three stakeholder archetypes: the reluctant data contributor, the unsupported activist, and the well-intentioned practitioner. Each archetype has its own role to play in realizing a participatory future. For example, data contributors are often victims of non-consensual AI practices. Meanwhile, practitioners take on a brokerage role of trying to create meaningful participation. In sum, we contribute a general theoretical framework of meaningful participation, empirical evidence of how secondary stakeholders navigate this framework, and a call to action for deeper interrogations of AI stakeholder needs.
\section{Related Work}
\label{sec:rw}

\subsection{Stakeholdership in AI} 
In 1984, Freeman introduced the term \textit{stakeholder} in organizational sciences as a new way of framing ideal corporate strategy for firms~\cite{Freeman2010-ww}. Stakeholder theory asserts that the firm is a network of value-creation obligations for those who can affect or be affected by the realization of an organization’s purpose (the wide definition) or those without whose support the organization would not exist (the narrow definition)~\cite{Dmytriyev2021-lj}. Thinking about stakeholdership has spurred numerous taxonomies and frameworks to consider both inclusion and prioritization. For example, a common taxonomy considers the stakeholder's degree of closeness to the firm's impact; primary stakeholders are directly affected, while secondary stakeholders have less clear relationships with the firm~\cite{Clarkson1995-we}. In participatory design, the practitioner is tasked with identifying a group of ''key'' stakeholders whose concerns should be prioritized during the design process~\cite{Friedman1996-kx}. Recent HCI work has focused on using stakeholder prioritization as a mechanism to empower marginalized communities. For example, critical race theory for HCI argues that vulnerable or systemically marginalized groups must be considered as key stakeholders~\cite{Ogbonnaya-Ogburu2020-xw}. Though the concept originated in organizational sciences, stakeholders are a crucial concept in design.

In the development of AI/ML tools and systems, stakeholder theory has been heavily adapted to inform who should be considered during the design process. The idea here is that the AI system is the firm and has a set of value-creation obligations to others. Similar to organizational research, the term stakeholder carries moral weight; when we fail to include stakeholders in the design process, we are skirting a moral obligation to do right by them. This justice-oriented approach to stakeholder inclusion is common in many human-centered machine learning approaches~\cite{Zhu2018-ov, Neumann2022-th, Chancellor2023-gd}. In fact, many works focus on the material harm caused by flawed stakeholder reasoning in AI and ML systems. For example,~\citet{Stapleton2022-bu} found that including parents in predictive child-welfare algorithm design fosters a richer understanding of the algorithm's potential shortfalls. \citet{Chancellor2019-tv} found that predictive mental health researchers have not coalesced on a conceptualization of stakeholders. This discord can cause dramatic differences in data handling procedures and limit human agency in predictive systems. 

\edit{An additional challenge in adapting stakeholder theory to AI/ML systems is the question of scale. In design, stakeholder inclusion is often proposed for small groups or community representatives. For example, Value-Sensitive Design (VSD) articulates indirect stakeholders as the large population of people who can be affected by a system without interacting with it~\cite{Friedman1996-kx}. Recent critiques highlight how this breadth of stakeholdership may neglect considerations of marginalization~\cite{Ogbonnaya-Ogburu2020-xw, Davis2015-fj}. Additionally, transitioning from individual needs to large stakeholder groups complicates engagement~\cite{Suresh2024-si, Dantec2013-wu}. When current AI systems are public-facing, is everyone a stakeholder? Engaging with AI stakeholders demands that we simultaneously consider contextual stakeholder positionalities while being able to abstract into shared ethical norms.} One possible solution is to encourage AI practitioners to engage with more rigorous stakeholder analyses. A large thread of AI ethics work involves identifying relevant stakeholders and stakeholder needs for specific problems such as interpretability~\cite{Suresh2021-lg} and explainability~\cite{Preece2018-qc}. Inspired by this work, our paper describes the specific ways that certain stakeholder groups realize participatory AI ideals.

\subsection{Participation in AI}
Distinct from ``AI Stakeholders,'' a similar thread of work focuses on ``AI Participation.'' While \textit{stakeholder} originates from organizational sciences, \textit{participant} originates from design~\cite{muller1993participatory}. Here, participants are the folks who we engage with during participatory design practices~\cite{Sanders2002-tn, Kensing1998-gk}. Given these origins, work on participatory AI---a genus of participatory design focused on AI tools---often focuses on the \textit{how} moreso than the \textit{who}.

In AI ethics research, there is an overwhelmingly positive vision of \textit{meaningfully participating with stakeholders}. Defining meaningful here has been a difficult research problem, dating back to civic engagement~\cite{Arnstein1969-bb, Fung2001-vt}. In AI contexts, these frameworks inform how we give technological ownership to---in other words, participate with---a specific stakeholder group related to the project (e.g., end-users~\cite{Bird2022-yu}, community organizers~\cite{Lee2019-dk, Suresh2022-nd}, domain experts~\cite{Smith2020-mq, Lee2020-al, Holten-Moller2020-fx}). Many of these creative approaches are successful at engaging with their intended participants.   

In addition to the successes of meaningful participation, recent work has explored the harms of superficial participatory AI practices. Often dubbed as ``participation washing,'' there is a class of stakeholder engagement approaches that do not constitute meaningful participation~\cite{Sloane2022-va}. For example, \citet{Corbett2023-yi}'s review of PAI projects found that numerous papers were attempting to ``fix'' participant opinions rather than engage with them. \citet{Delgado2023-bo} describe a similar spectrum of consulting-based participation to ownership-based. To these forms of non-meaningful participation, recent work has turned its gaze toward practitioners themselves. Unpacking meaningful participation involves reflecting on the contextual power dynamics, such as the relationship between a researcher and their participants~\cite{Vines2013-gj, Frauenberger2015-nn, Bondi2021-lp}. Indeed, there is a current seed in participatory AI that is starting to consider the stakeholders that have influence beyond the primary group of participants. 

In this paper, we build on this seed by focusing on stakeholders who are not obvious participants, such as data contributors, activists, and practitioners. Inspired by work that calls for us to do right by these stakeholder groups~\cite{Sloane2022-va, Li2023-wa, Moller2020-rf, Sambasivan2021-zw}, despite their indirect relationships to AI, we consider the ideals of meaningful participation for secondary stakeholders. To do so, we must acknowledge that broadening who we consider complicates the participatory ideals established in prior work. What do ownership and agency look like when we start to scale stakeholder considerations?

\section{Methods}
\label{sec:methods}
Our research focuses on secondary stakeholders' experiences with AI participation. Therefore, we conducted semi-structured interviews with 12 individuals who are researchers, developers, activists, or data workers. We chose to interview people in these roles as previous work has heavily theorized how they influence AI systems~\cite{Li2023-wa, Pierre2021-zz, Suresh2022-nd}. We used an interpretivist methodology~\cite{Doyle2007-fh}  to capture our participant's understandings of their lived experiences and frustrations. Therefore, we (1) ran semi-structured interviews with open-ended questions and member checks, (2) analyzed interviews with inductive open coding~\cite{McGrath2019-co}, and (3) communicated with participants as needed.

\subsection{Participant Recruitment}
We circulated our call for participation in relevant venues such as researcher Slack groups, advocacy mailing lists, and moderator forums. Interested individuals completed an intake survey to check they met our eligibility requirements: (1) were 18 years or older and (2) had participated as an AI researcher, developer, advocate, or data contributor over the past 24 months. We selected participants using purposeful stratified sampling techniques~\cite{Onwuegbuzie2007-qx, Robinson2014-us}. Because we are interested in comparing and contrasting participation groups (researchers, developers, activists, and data workers), we wanted to ensure adequate representation and expertise in all groups. Our participants include international and US-based interviewees. Our study was approved by the university Institutional Review Board (IRB) \#00020382. Given some of our participants are members of adjacent research communities and could easily be identified, we present our participant demographics in aggregate (Table~\ref{tab:demographics}). Table~\ref{tab:psuedonyms} details the participants' pseudonyms and their role descriptions. We chose to use pseudonyms instead of numeric identifiers to communicate the personal nature of our participants' stories. Company and institution names have been redacted unless the participant explicitly requested otherwise. Note that many participants embody more than one secondary stakeholder role. For example, Katie became an activist after her frustrating tenure as a data worker.

\subsection{Interview Procedure}
Our interview questions were designed to understand (1) the participant’s experience with AI, (2) their frustrations with their current role in AI practices, and (2) their personal vision for a future of AI participation. Participants completed a written consent form before scheduling their interview. At the beginning of each interview, the interviewer verbally confirmed participant consent and reminded the interviewee they can decline any questions or redact any information after the interview. As the researchers may have personal beliefs and biases on AI, we followed \citet{Doyle2007-fh}'s work on consistent member checking to make sure the participants' responses are interpreted correctly. For instance, by asking questions such as, \textit{``So I'm hearing you say XYZ. Is that a fair interpretation?''} and encouraging participants to correct the interviewer's interpretations. We also validated our interpretations by communicating with participants after interviews as needed. The interviews were conducted virtually through Zoom and lasted approximately $1$ hour. All participants were compensated $\$25$ USD in gift cards for their participation.

We adopted an inductive approach~\cite{thomas2003general} to our analysis inspired by interpretivist hermeneutic methodologies~\cite{Doyle2007-fh}---techniques to understand how people negotiate the meaning of a specific term (i.e., AI participant). Given how interview participants' relationships with AI are central to our research question, we employed a stratified version of open coding. We began with iterative open coding of each individual interview. Two researchers were involved in the analysis process. Next, the two coding authors transferred all of the codes into a virtual whiteboard coloring each code based on the participant's role (i.e. researcher, data worker, etc). This coloring allowed us to compare and contrast themes across groups. Finally, all authors contributed to the synthesis of major themes presented in this paper. Since synthesis in thematic analysis runs into the risk of diluting participant experiences into the author's narrative~\cite{Braun2006-eg}, we engaged in peer debriefing~\cite{Lincoln1985-qf} among authors to shed light on our positionalities. We reached saturation after an initial round of interviews with three participants from each positionality (n=12). This sample size meets the requirements of interpretivist and constructivist methodologies~\cite{Soden2024-cf, Onwuegbuzie2007-qx, Doyle2007-fh, Cunningham2017-bq}, particularly given our focus on individual positions about AI participation. 

\section{The Meaningful Participation Ladder}
\begin{figure}
    \includegraphics[scale=.15]{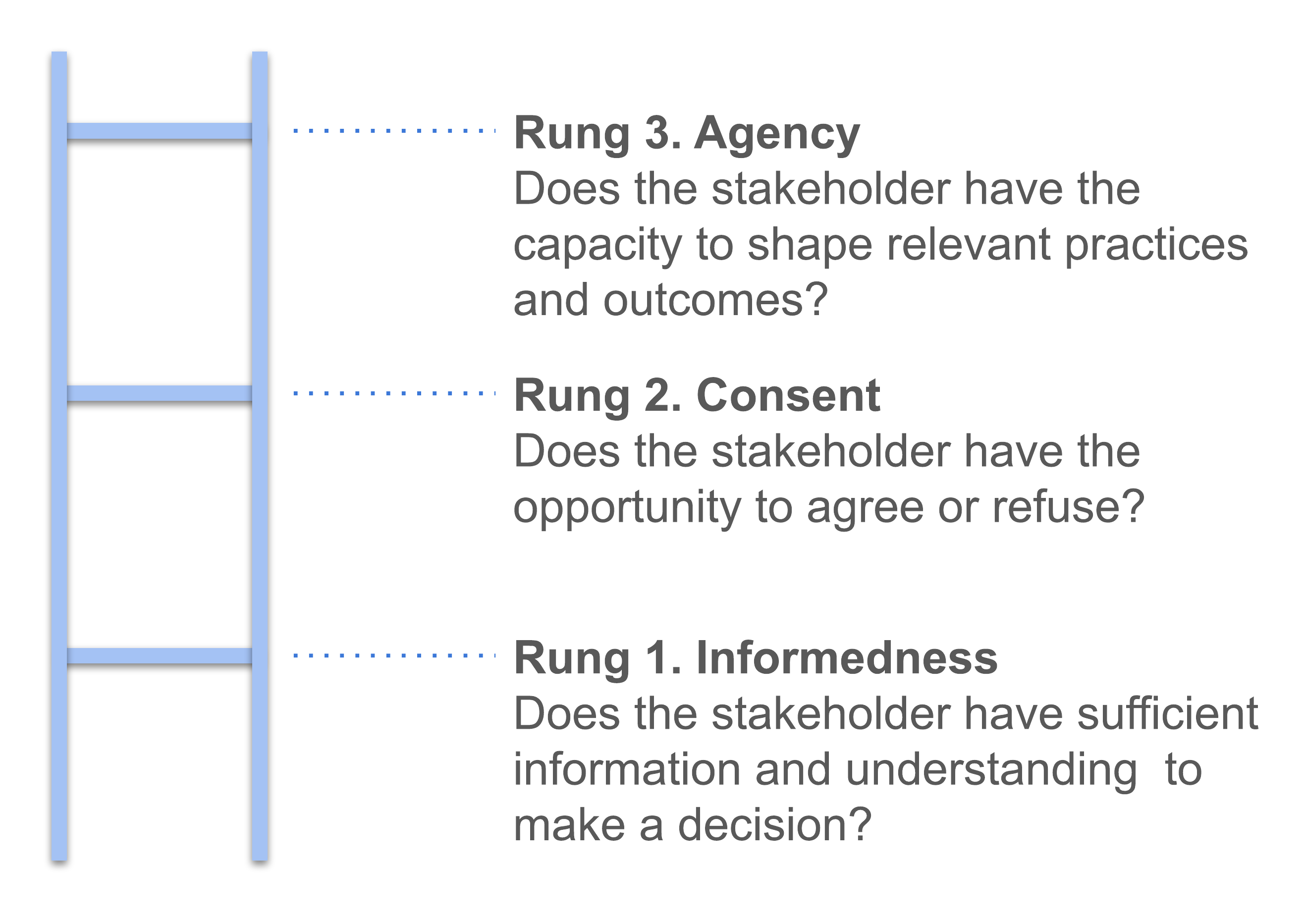}
    \Description{Four-stage diagram overviewing our vision for participatory AI. First step is a ladder with rungs of informedness, consent, and agency. Second stage is definitions of each term. Third stage is actions for researchers. Fourth stage is actions for policymakers.}
    \caption{Overview of how we envision a ladder of meaningful participation. Each rung of the ladder is a participatory ideal that we argue must be realized for both primary and secondary AI stakeholders.}
    \label{fig:ladder}
\end{figure}

We start by articulating a general framework of meaningful participation that applies to both primary and secondary stakeholders. One area that has deeply considered ``meaningful'' participation is civic engagement. Specifically, Sherry \citet{Arnstein1969-bb} proposed the ladder of citizen participation, which models the degree to which civic processes redistribute power to citizens. Recent work in AI ethics has focused on translating this idea of the ladder of participation levels to build normative frameworks for participatory AI approaches~\cite{Suresh2024-si, Corbett2023-yi, Delgado2023-bo}. We seek to marry this work with current AI ethics calls to create a general framework for considering meaningful participation. Given that our normative goal is to broadly do right by people, we translate~\citet{Ajmani2024-qk}'s recent FAccT work on justice in AI to the ladder of participation. Specifically, we start from their position that \textit{``agency is a contributor to justice and a product of consent policies.''} In this section, we synthesize the work of \citet{Arnstein1969-bb}, \citet{Ajmani2024-qk}, and their peers~\cite{Cooper2024-xg, Chancellor2023-gd, Birhane2022-nd} to describe meaningful AI participation. Meaningful participation is the step-by-step process of building (1) informed, (2) consensual, and (3) agentic relationships with AI, where you cannot reach a higher rung without establishing the lower rungs first (Figure~\ref{fig:ladder}). In later sections, we use this ladder to describe each secondary stakeholders' role in AI systems and their complications in realizing these three participatory ideals.

\textbf{Rung 1. Forms of Informedness:} We define an informed stakeholder as an individual who has sufficient information and understanding at hand to make a decision about their participation~\cite{Salmons2017-yo, Office-for-Human-Research-Protections-OHRP-2010-nv, Henderson2011-ns}. Informedness of both primary and secondary stakeholders is a perennial challenge in AI research and development due to the scale of training data needed~\cite{Chancellor2019-cz}, the technical nature of AI development~\cite{Zhang2023-ta}, and the fear of new technology~\cite{Ghoshal2020-xz}. Expertise gaps exacerbate peoples' ability to be informed -- it is easier to make informed judgments about AI participation when you have relevant technical knowledge, which most people do not have. In settings with clear primary stakeholders, such as PAI research projects with human subjects, previous work has called for minimizing these expertise gaps between researchers and participants~\cite{Zhang2023-ta, Birhane2022-nd, Bondi2021-lp}. This work underscores how AI participation relies heavily on the informedness of stakeholders. Often, practitioners are responsible for informing the stakeholders they directly interact with~\cite{Cooper2024-xg}. However, when considering secondary stakeholders, there is minimal opportunity for interaction. In later sections, we describe how, though informedness is the most basic level of cultivating meaningful participation, it is extremely difficult to realize at the scale needed for secondary stakeholders.

\textbf{Rung 2. Consent Builds on Informedness:} Consent is about giving stakeholders the opportunity to agree to or refuse participation, free from coercive forces~\cite{Friedman2019-aw}. Consent scholars generally agree that uninformed consent is not true consent~\cite{Beres2007-is}; informedness and consent go hand-in-hand. For example, recent work evaluating common data consent procedures, like Terms of Service agreements, demonstrated that such broad agreements cannot feasibly create a consentful environment~\cite{Ajmani2024-qk}. Often, these are our only mechanisms for soliciting consent from secondary stakeholder groups, such as data contributors. While there are repeated calls for legible~\cite{Luger2013-do}, informed~\cite{Utz2019-ay}, and affirmative~\cite{Im2021-uu} consent procedures online, they have not come to fruition. Building on previous work on ethical tensions in AI/ML~\cite{Chancellor2019-cz}, consent in AI contexts must happen at a massive scale. Researchers and developers have techniques to solicit consent from primary stakeholders in PAI research projects~\cite{Office-for-Human-Research-Protections-OHRP-2010-nv}, but it is unclear if these can properly scale. While consent for secondary stakeholders is a necessary step for meaningful participation, it is not an easy task. In this paper, we describe the tensions of getting consent-at-scale from well-intentioned AI practitioners and the frustrations of non-consensually engaging with AI.

\textbf{Rung 3. Agency Beyond Consent:} Even in ideal worlds with informed and consenting stakeholders, meaningful participation involves agency. In PAI, agency is about giving participants the capacity the shape the outcomes of their participation, such as resulting datasets~\cite{Kuo2024-nx}, tools~\cite{Smith2020-mq}, and algorithmic recommendations~\cite{Stapleton2022-bu}. Moreover, agency captures one's capacity to shape their own participation; agency involves allowing participants to voice their opinions freely. For an example of anti-agentic but informed and consensual PAI practices, we can look to \citet{Corbett2023-yi}'s examples of PAI projects where researchers tried to ``fix'' participant opinions rather than engage with them. On a broader scale that considers secondary stakeholders, \citet{Ajmani2024-qk} articulate data agency as one's capacity to shape action around their data, concluding that non-consensual data practices are anti-agentic and, more seriously, unjust.


In sum, meaningful participation involves designing for three participation ideals: informedness, consent, and agency. However, these are big and broad asks when engaging with secondary participants. Standard PAI techniques of achieving informedness and consent with primary stakeholders start to break down. Lending agency to AI stakeholders must happen in a scalable way. In the rest of this paper, we use this ladder as a framework to better understand the empirical experiences secondary stakeholders have with AI.

\section{Secondary Stakeholder Archetypes}
\label{sec:archetypes}
In this section, we present three archetypes of secondary AI stakeholders and how they traverse the meaningful participation ladder. \edit{The term ``stakeholder'' is often broadly and ambiguously defined in HCAI~\cite{Chancellor2019-tv, Delgado2023-bo}. We present the experiences of a specific class of ``murky'' AI stakeholders: those with complicated relationships to AI systems they affect. We find this aligns with the concept of secondary stakeholders from organizational sciences~\cite{Clarkson1995-we}: those (1) influenced by a firm’s decisions but (2) outside formal contracts. However, we find that this definition is complicated by the ambiguity of AI. For instance, one participant, Fiona, had a contract as a paid moderator, but did not necessarily have a clear contract with any AI system, as she did not realize her work was training one. We use \textit{secondary stakeholder} to describe individuals who (1) influence AI systems broadly but (2) lack a direct contract with an AI system. While this working definition emerged from our analysis, participants occupied nuanced positions, elaborated below.}

To start, we introduce the forms of AI participation our interviewees described. AI participation is split between human-centered activities that are grounded in traditional research methods~\cite{Stapleton2022-bu, Kuo2024-nx, Zhang2023-ta} and data-centered activities. We find evidence that secondary stakeholders are delegitimized in their participation with AI systems. 

To relegitimize their participation in AI systems, we frame stakeholders' contributions into three stakeholder archetypes: the reluctant data contributor, the unsupported activist, and the well-intentioned practitioner. For example, data contributors are often victims of non-consensual AI practices. Meanwhile, practitioners take on a brokerage role of trying to create meaningful participation. By conceptualizing each archetype along our ladder of meaningful participation, we identify where current gaps are in PAI practices. Each archetype has its own role to play and actions they take in realizing a participatory AI future.
\subsection{Forms of Participation}
\textbf{Human-centered participation }in AI research and development is a well-defined set of methods where ``AI participants'' are the people actively engaging in an activity (e.g., workshop, co-design, prototype evaluation~\cite{Sanders2002-tn}). These participatory efforts often engage with primary stakeholders, such as end-users or those directly affected by the AI system~\cite{Smith2020-mq, Stapleton2022-bu, Lee2020-al}. Our interviewees noted participating in AI-related development efforts, such as design workshops or research studies, as a clear way of defining ``AI participants.'' Maddie, a non-profit industry researcher, initially described participation as research-based: \textit{``When I hear the term participant, I'm immediately thinking about participants in our studies, the way that I'm a participant in your study.''}

As participatory methods have evolved to be more collaborative between researchers and participants, our interviewees disagreed on whether practitioners---individuals who facilitate participatory AI workshops and activities---were participants themselves. This murky position of practitioners highlights their role as secondary stakeholders. For example, Sean is a researcher who uses participatory AI methods to build agriculture technology. He self-identified as an AI participant insofar as he participates with farmers (i.e., his research participants) to build a usable tool. In that way, his ideas and values bear on the final product. In contrast, Tyler is a researcher using participatory AI methods to build tools for online communities, such as Wikipedia. He noted that researchers are not inherently participants in the AI systems they are building. However, depending on the context, researchers can become members of the communities they study. For example, Tyler has become an avid Wikipedia editor on the side of his Wikipedia-based research. We find that human-centered participation has a clear set of primary stakeholders -- the people who are engaging in the participatory design activity. Meanwhile, practitioners operate as secondary stakeholders; \edit{Tyler uses PAI methods as a researcher, but does not have a formal contract with any specific AI system. We find that this distance influences his perception of his own role as a participant, raising an open question of how researchers relate to “participatory AI” broadly.}

\textbf{Data-centered participation} deviates from the active type of participation found in participatory design methods~\cite{Delgado2023-bo}.  Data-centered participation relies heavily on secondary stakeholder groups, such as data contributors and AI system developers. Yet, our interviewees cited data contribution and moderation as forms of AI participation. Building on previous literature that highlights how data work can be participation~\cite{Sloane2022-va, Li2023-wa}, we find that individuals can participate in AI by contributing work that helps build AI training datasets.

Data-centered participation can include almost everyone who uses the internet. Our interviewees reflected on how they passively participate through data in their everyday activities. Scott develops AI tools for other developers, and he noted: \textit{``I see participating in AI as contributing to the tech or contributing to the algorithms. For products like [AI~tool], I do both.}''
Scott described building AI tools as ``active'' while contributing his code as training data is ``passive,'' because he simply makes his code publicly available online. Many of our interviewees described similar situations; they produce data used for AI training data simply by engaging in adjacent activities, such as development, writing, and publishing online content. In terms of stakeholdership, this passive data contribution \edit{is secondary, as it exists outside of any formal contract}. Yet, our interviewees described their own frustrations with feeling like everything they publicly create is unintentional training data. We elaborate on these frustrations in later sections.

Meanwhile, many of our interviewees similarly used the term ``active'' to describe their roles as intentional data workers. Data worker is the umbrella term used to describe participants who actively contribute, annotate, or moderate data that is known to be used for AI model training~\cite{Li2023-wa}. This includes volunteer work, such as contributing to Wikipedia, and paid positions, such as being an expert annotator\footnote{We note that these roles (as data participants in an employment relationship vs. volunteering) exist in different legal contexts.} For example, employers often own the outcomes of their employees' work. As a positive example, Falen works as an expert data annotator for a mental-health-based AI company and self-identified as an AI participant. She was hired to train a model with her clinical mental health expertise. Falen positively noted: \textit{``I'm helping train the machines so that they can help therapists do really important work.''} However, Falen represents a rare case. Overwhelmingly, we find that data workers have little information, opportunities to consent, and agency. 

In the following sections, we highlight how the scale of data-centered participation complicates the stakeholder considerations of human-centered participation. From data workers feeling exploited to practitioners feeling like their efforts are hollow, realizing participatory ideals is extremely frustrating for everyone.

\subsection{The Reluctant Data Contributor: Navigating Informedness, Consent \& Agency}
In this section, we introduce \textbf{the reluctant data contributor}, who---if given sufficient information and the opportunity to consent to their current AI participation---would actively choose to withdraw their data from AI training datasets. This archetype contrasts Falen, who is participating in AI willingly and positively. It is important to note that none of our participants described reluctant participation for human-centered efforts (like deliberate inclusion in research). Our interviewees felt reluctant about their data and data work being used to create training datasets for AI models.

\subsubsection{Informedness. }
Our interviewees described how many contributors in data-centered participation are uninformed and, subsequently, would not agree to their data being used for AI training if they were informed. The most straightforward example of an informed but reluctant data contributor came from Sabrina, who described her complicated situation as a volunteer moderator. She moderates a platform that is a clear data source for LLMs. Given her other position as a researcher, Sabrina understood her data work was being used to train LLMs but was powerless to stop it. Sabrina noted her own unease in being an AI participant, using the term reluctance: \textit{``It's something that I know, but I'm a reluctant participant in this.''}

Sabrina's experience highlights how volunteer data workers, who are often crucial moderators for their online communities, are aware of how their data is used but are unwilling. At the same time, Sabrina described how many of her moderator colleagues are uninformed. \textit{``I understand the basics whereas it’s not something that [other moderators] are thinking about because it’s just like if you know, you know.''} Sabrina speaks to an important point -- though many of our interviewees were informed, given their expertise, many other data workers are often uninformed.

Further complicating informedness, data-centered participation can be passive. Therefore, a myriad of internet users are uninformed data contributors. For example, Maddie described her realization that all of her publicly available content, such as her research blogs, is now training data:
\begin{quote}
   \textit{``Most of that data is made by unknowing, perhaps unwilling participants. Who wrote the research blogs? Who wrote the books? Who wrote the Reddit posts? The Wikipedia articles? Their data is now training AI.''}
\end{quote}
Maddie empathized with this feeling of unwillingness embraced by Sabrina. Maddie continued to describe that, even though she is aware her data is probably being used for AI training, she had no say in the matter.

\subsubsection{Consent.}
Both Maddie and Sabrina represent informed but non-consenting participants. While informedness describes a participant's level of understanding, consent describes whether they have actively agreed to their participation. In data-centered participation, our interviewees described a significant lack of opportunity for them to consent without leaving the platform altogether. Sabrina explicitly called out that her continued data work should not be seen as consent.
\begin{quote}
    \textit{``It’s like you can opt out of the community, or you can be relinquished to having your data used. Those are the only two options. There’s no way to give back to your individual community, which means a lot to you.''}
\end{quote}
In this case, Sabrina's only recourse is to withdraw her data work entirely rather than simply withdraw from downstream AI uses. In other words, her engagement with her online community carries AI participation with it. 

Our interviewees described how, even when they have consented, current data practices often take this agreement out of scope. For example, Fiona worked as a commercial content moderator and, therefore, had a contract where she consented to her data contribution being used by a company. However, she articulated that her consent was taken out of scope when she was forced to moderate gruesome content,
\begin{quote}
    \textit{``When they did the screening and the interviews, they just told us that the job might be slightly challenging. They even showed us samples...but the content that we ended up moderating was directly from the war in [home country]. Everything was gruesome videos. It's the worst thing that one can imagine.''}    
\end{quote}
Fiona felt informed at the time of consent but was then given no recourse to withdraw her consent in certain contexts. Contrast this to Falen's willing and consentful participation. Falen described how, given she annotates severe mental health content, her company offers the ability to refuse certain data annotation tasks: \textit{``We can say this session is too hard to code...And they give us another one that's not as emotionally taxing.''}
Falen's experience offers a positive future where data contributors have multiple opportunities to agree to AI-training tasks and refuse tasks where contextually appropriate---turning \textit{reluctant data contribution} into \textit{willing data contribution}.

\subsubsection{Agency.}
We conceptualize the reluctant data contributor as someone who has little to no agency over their data being currently used to train AI systems. Often, the only recourse these contributors have is dramatic non-use. This lack of agency is represented in Sabrina's feeling of voicelessness: \textit{``You almost don’t want to admit that your work is being exploited, but you don’t have any sort of voice.''}

Katie, a former data annotator on mTurk, described feeling like she had agency only after she reckoned with the task of annotating a politically charged dataset:
\begin{quote}
    \textit{``I wish when I started this work that I had been more aware. I didn't realize at the time that what I was doing could be used for something that I personally didn't agree with. It was honestly a task tagging border crossing photos that stopped me, and I thought: wait a minute.''}
\end{quote}
In this situation, Katie did have the agency to choose whether to complete this annotation task or not. However, her lack of informedness clouded this choice, highlighting how agency builds on informedness. Since then, Katie has become a data rights activist to advocate for better practices around soliciting data annotation. Note that if the reluctant data contributor did have sufficient agency, much like Katie chose to stop her data annotation task, they would choose not to contribute their data.

\subsection{The AI Activist: Fighting for Informedness, Consent \& Agency}
We interviewed multiple people with similar stories to Katie: they became activists after their own frustrating experiences of being AI-oriented workers. In this section, we introduce the archetype of the \textbf{AI activist.} We describe this stakeholder as someone who fights for informedness, consent, and agency on behalf of both primary and secondary stakeholders. We find that activists fight for more meaningful participation by advocating for better general AI initiatives, such as AI literacy, and better specific practices from platforms, developers, and researchers.

\subsubsection{Fighting for Informedness.} \label{fighting_for_informedness} We find that activists are currently fighting for informedness on two fronts: (1) public informedness on AI and (2) practitioner informedness on current practices. In terms of fighting for public informedness, Kayla described her advocacy work on AI literacy. She described AI literacy as a balancing act between informing the public, but also reducing the harm caused to those who do not understand the technical details of AI,
\begin{quote}
   \textit{ ``I'm incredibly in favor of AI literacy... but with what other product do we ask people to have to be that literate? We don't expect ourselves to be fully literate about our kitchen appliances. We expect that regulators, government officials, and companies, are responsible for making sure the products we use are safe and fair and and ethical to some degree. To some extent, does this push towards AI literacy just displace that responsibility onto the individual.''}
\end{quote}
Kayla's fight is to consider \textit{how} to inform participants and \textit{who} is responsible when participants are left in the dark.

At the same time, Katie described much of her activism as fighting for more transparency about current practices. For example, she described the invisibility of data work in current AI practices,
\begin{quote}
    \textit{``The tech companies not wanting to acknowledge that there is this workforce is very frustrating. Data workers put in essential work. These datasets have been built on the backs of data workers and to keep them invisible is disrespectful. }
\end{quote}
In this case, informedness and transparency go hand-in-hand. Katie described her ideal world as one where data workers are informed about the tasks they are completing, and practitioners are informed about the labor behind data sources.

It is important to note that, especially in Katie's case, fighting for informedness requires multi-stakeholder support from researchers and developers. For example, both Katie and Fiona described their positive experiences of going to research conferences to tell their stories.

\subsubsection{Fighting for Better Consent Procedures.}
Our interviewees described fighting for better consent procedures as a process of navigating the real-world limitations of current consent practices. For example, Katie described how the current consent mechanism on mTurk is impractical for data workers where ``time is money:''
\begin{quote}
   \textit{``For some tasks, the consent documents are very long. And I know that most workers skip them because time is money. They just hit the okay button and go."}  
\end{quote}
Therefore, Katie is fighting for concise consent documents that contain the necessary information about the task, including the intended data use. Kayla similarly described thinking about consent as context-specific rather than as a one-size-fits-all best model. Conceptually, Kayla described her fight as moving away from any singular consent model:
\begin{quote}
    \textit{``Rather than thinking about a single consent model, what happens if we think in terms of coercion or coerciveness? Thinking about where people are giving up power because they have no other choice''}
\end{quote}
Through her advocacy work, Kayla builds tools and frameworks for evaluating points of coercion in current data practices. To do so, Kayla advocates for conceptualizing coercion as an opposing force to consent. Inspired by feminist work, she repeated the mantra, \textit{``consent with coercion is not consent.''}

Katie and Kayla highlight how fighting for better consent procedures happens on two fronts. First, the pragmatic fight of figuring out appropriate consent procedures for a specific AI context -- such as mTurk workers. Second, the conceptual fight of building more robust ways to evaluate current consent practices. Addressing both of these concerns is a crucial step toward agency, as agency can be communicated and afforded through consent.

\subsubsection{Fighting for Agency.} \label{fighting_for_agency}
A lack of agency in AI relationships is a major motivating factor for AI activists. Kayla described how considering AI participation without considering agency can lead to non-meaningful participation,
\begin{quote}
    \textit{``I think sometimes participation gets confused with meaningful ownership or control over technologies so we can have participants without agency.''} 
\end{quote}
In other words, Kayla notes that using the term ``participation'' without creating space for ownership creates a problematic class of AI participants who do not have any say in how their participation affects the system. In an ideal world, this is considered unethical. Therefore, Kayla viewed advocating for agency and advocating for ethical AI as one in the same.

More concretely, Fiona is advocating for agency for her specific community of refugee moderators who were tasked with moderating violent content. Given the extreme consequences of Fiona's participation, her activism work involves community building and collective action. For example, she described creating a support group for her fellow moderators, a reparative measure from the isolation she experienced as a moderator.
\begin{quote}
    \textit{``We were never given a chance to talk about them to anyone...We're having nightmares. We're having insomnia. We're having suicidal thoughts.''}
\end{quote}
Fiona views fighting for agency as fighting for support of past moderators and regulation for future moderators. Given that agency is one's capacity to shape action, Fiona argues that support and community are two crucial factors in lending agency to large stakeholder groups, such as commercial moderators.

Our activists described their own struggles with agency and resistance-based efforts. Positively, Kayla described an inspirational history of resistance-based design as a mechanism for groups to reclaim agency. 
\begin{quote}
    \textit{``There's a really fascinating counter-history of design, which is how people express and create tools that work for them at times when like a lot of their choice or ability to live freely has been foreclosed.''}
\end{quote}
These acts of reclaiming agency, power, and design are effortful. Kayla's conceptualization of agency as being effortful to gain was mirrored in Katie's experience as a data worker. Katie described the transition from data worker to data rights activist as a necessary step she needed to take to gain agency: \textit{``It does take a lot of time and energy to advocate for data rights. You're going up, and you're speaking out.''} While resistance-based efforts are necessary to create change, to what end are we putting AI activists into unsupported spaces and expecting them to speak up while they are in already burdened positions? We find that activists need multi-stakeholder support from researchers and designers to make meaningful change.

\subsection{The Well-Intentioned Practitioner: Brokering Informedness, Consent \& Agency}
For this archetype, we focus on \textbf{well-intentioned practitioners}, AI researchers and developers who are truly attempting to engage in best practices for participatory AI. Although some practitioners, such as Tyler, did not view themselves as participants, they heavily engaged and participated with communities and recognized their own influence on AI development - effectively working as brokers of informedness, consent, and agency to other stakeholders. However, these well-intentioned practitioners also recognized the limits of brokering these values because of scale, which limited an ideal participatory future with AI.

\subsubsection{Brokering Informedness. } Our interviewees described successfully brokering informed AI relationships with primary stakeholder groups such as research participants and end-users. Practitioners often brokered informedness at the time of consent as an essential part of participatory AI -- a required practice in human subjects research. Tyler described informing research participants as the process of: \textit{``Giving people enough information to articulate their ideas and make the right decision.''} Here, informedness is a necessary pillar in one's decision-making for research participants. Maddie noted that informedness in research settings is often risk-centered:
\begin{quote}
\textit{``We try to make [consent documents] very simple, but not in the way of removing important details. We try and make it as intuitive and readable as possible. So people understand the risks of what they're doing.''}
\end{quote}
We find that practitioners who use participatory AI methods are trying to create legible documents to achieve participant informedness during consent. However, given AI technology is relatively new and complex, our participants told us their participants are nervous because of the unknown outcomes of AI systems. Therefore, AI researchers and developers have to build participant relationships and understanding to supplement informedness with trust to move forward. 

For example, Sean---a researcher who uses participatory AI methods---struggled to even introduce the idea of AI as a solution to his research participants. When working on AI tools for farmers, Sean mentioned:``\textit{Trying to bring new ideas to those farmers was a major challenge because they fear that it will disrupt their way.}'' To help manage this unease, Sean brought in agricultural extension workers as mediators between himself (the researcher) and the farmers. This relationship-building and brokering is feasible in well-defined research settings and served Sean well as he could better engage with the community of research participants. 

However, as AI participation has expanded beyond well-defined research settings, informedness has become a widespread issue. Maddie often uses public datasets for training models, and described how she tries to ``inform'' a different stakeholder -- data contributors. To reach this group, she described her research on blogs and social media, something she felt obligated to do:
\begin{quote}
    \textit{``I feel it's my duty as a researcher to engage at least a little bit, so I write Medium articles about things people should know about large language models as a user.''}
\end{quote}
However, Maddie recognized that these broader efforts can feel ``hollow''; she explained that there is currently no way to guarantee that these messages are reaching the people whose data she is using. We find that practitioners are well-equipped and often willing to broker informedness in human-centered participation, where they can converse with their research participants, but are unable to realize this ideal in data-centered participation.

\subsubsection{Brokering Consent. }
We find that practitioners have similar difficulty in brokering consent -- current practices don't scale well beyond human-centered participation. Many practitioners described their practices of creating consent documents, reminding AI participants they can withdraw at any time, and even reaffirming consent after engaging with participants. These are best practices in research settings. 

However, similar to brokering informedness, standard consent practices break down with data-centered participation. Charles, a developer who uses a variety of AI-related methods, noted the difficulty of getting consent in a space with no policy guidance:
\begin{quote}
\textit{ ``A lot of AI projects do not require consent because it's this new space. People still don't know how to navigate it or how to regulate it, so a lot of people are building models without consent.''}   
\end{quote}
Maddie also described her personal struggles with navigating consent in non-human subjects research. While she believes in individual informed consent for data usage, she had trouble operationalizing it for her research datasets because of their size:
\begin{quote}
    \textit{``You cannot get individual consent for every single person. I mean, you could if platforms were better. They could have each person opting in or out.''}
\end{quote}
Maddie continued to speculate about how platforms could get consent for all data-centered contributions, referencing previous work that articulates consent-at-scale as a difficult and unsovled problem~\cite{Barocas2014-in}.

Sabrina---in part due to her position as both a researcher and a moderator---believes this is a solved problem, and very clearly placed the onus on platforms and platform design. She described the problematic lack of incentive structures for platforms and companies to design more responsible consent mechanisms.
\begin{quote}
    \textit{``We know that it's possible for people to consent and to have this. But [companies] are not doing it, and they're not going to. They don't have to.''} 
\end{quote}
Sabrina is referencing the lack of laws around obtaining meaningful data consent. Sabrina brought up that there have been promising movements for better consent procedures, such as the AI Bill of Rights~\cite{The-White-House2022-se}, but there is little incentive for platforms to realize these calls into design. Moreover, there is no punishment for using problematic consent procedures, such as Terms of Service agreements.

\subsubsection{Brokering Agency. }Finally, we found practitioners heavily consider the agency of the AI participants they engage with. Our interviewees described agentic practices---practices that afford agency to participants---in their own disciplines. For example, when Sean was researching and developing AI tools for farmers, he consulted farmers at every major step of the process. 
\begin{quote}
    \textit{``We went to the field and collected the needs of the farmers before we eventually came back. We came up with our model and then gave it to the farmers to get feedback from their usage. Based on that feedback, we improved the model and eventually deployed it.''}
\end{quote}
Feedback solicitation and iteration are just one mechanism for giving participants agency in the AI development process. Anthony, an AI tool developer, described soliciting feedback on a larger scale with user surveys.
However, agentic practices take time. Even in well-defined and well-resourced environments, such as academic research and AI development, our interviewees wanted increased support for brokering meaningful relationships. For example, Spencer, who develops AI tools for the public sector, described the time-consuming process of building rapport and, thereby, more agentic relationships with his participants:
\begin{quote}
    \textit{``We are not delivering these public services, right?  We are not on the ground. We don't understand every possible constraint. A lot of time goes into understanding, having conversations with people who know the right answers, and making sure that the assumptions we make are valid.''}
\end{quote}
While this step of building understanding is crucial to ethical participatory AI practices, it is time-consuming. Therefore, Spencer proposed that researchers and developers get more support by having longer project timelines and increased resources so they can do right by the people who would be affected by their system. We find that practitioners are engaging in creative practices---such as rich feedback solicitation---to broker agency, but doing so can run counter to the fast-pace of current AI development.

\section{Discussion}
\label{sec:discussion}
\subsection{Recommendations for Meaningful Participation}
We return to the central question of PAI work posed earlier and one central to FAccT~\cite{Delgado2023-bo, Suresh2024-si, Cooper2024-xg}: \textit{how do we meaningfully participate with stakeholders?} \edit{Many of our interviewees described actionable ways to improve AI participation. For example, Maddie and Sabrina suggested default opt-in consent mechanisms. Spencer called for longer project timelines to incorporate participatory methods. In terms of the meaningful participation ladder, practitioners could broker agency with activists by inviting them to research venues, fostering a shared agenda. At the same time, we find that many of these recommendations rely on powerful stakeholders, such as technology companies, making them beyond the scope of any single archetype presented in our work.} 

Participating with secondary stakeholders has been heavily theorized in prior work~\cite{Sloane2022-va, Li2023-wa}, but is challenging due to scale and ambiguity about who to include and how to include them. Specifically, we align with \citet{Sloane2022-va}'s claim that all participation is work, and failure to change our PAI practice to include all stakeholders can lead to exploitative practices. In fact, to ignore AI labor is to ignore deep ethical and societal injustice~\cite{Li2023-wa}. Our empirical work and results show that the secondary stakeholders we interviewed are frustrated by current AI systems and their own anti-agentic relationships with them. Further, our participants highlighted that, without deep consideration of secondary stakeholders in PAI practice, we risk furthering exploitative practices of these groups rather than meaningful ones. 

\subsection{Using our Archetypes}
In addition to providing empirical evidence of risks with PAI and secondary stakeholders~\cite{Sloane2022-va,Li2023-wa}, we contribute several archetypes to the PAI space that focus attention and reasoning. We anticipate that many readers of this paper are one of our archetype---FAccT researchers may see themselves as well-intentioned practitioners. Like Maddie and Tyler, they are equipped to broker informedness, consent, and agency in human-centered participatory settings but feel limited in using the same strategies in data-centered settings, to the extent that Maddie's work feels futile. Using our archetypes may enable these practitioners to direct their efforts toward secondary stakeholders in a more nuanced and strategic way. For example, by applying the first rung of our ladder (i.e., informedness) to our most disempowered archetype (i.e., the reluctant data contributor), well-informed practitioners might consider how their brokerage could best inform data contributors to AI systems. 

\edit{Additionally, future research could use our archetypes to identify contextual user needs, akin to design personas. These contextual descriptions are crucial when operationalizing ideals like informedness, consent, and agency.} For example, Katie mentioned that being a data contributor meant often ignoring documentation, because \textit{``time is money''}. Our archetype articulates a stakeholder need in achieving informedness---to be informed in a way that is not labor-intensive; research with data contributors thus should consider the value of conciseness alongside legibility~\cite{Luger2013-do}, transparency~\cite{Barocas2009-vx}, and literacy~\cite{Bhat2024-je}. Similarly, our activist archetype requested support for multi-stakeholder action, highlighting a gap in current infrastructure to realize their positive visions. Like personas, our archetypes are both a descriptive synthesis of our participants' experiences and also can be generative for new design and technical PAI futures.  

\subsection{Recognizing Systemic Barriers}
It is important to note that not all of the onus is on researchers and practitioners to improve AI relationships with secondary stakeholders. \edit{While practitioners should engage with the above recommendations, our findings highlight the limits of achieving participatory AI. Our interviewees echoed the current calls in AI ethics literature for increased AI regulation and literacy campaigns~\cite{Undheim2023-jk, Ng2021-dc}; platform redesign for meaningful data consent~\cite{Im2021-uu, Ajmani2024-qk}; and social responsibility of tech companies~\cite{Camilleri2024-pg}. Practitioners hold privilege but face systemic barriers, such as tech companies failing to design better consent mechanisms. Our findings stress these real constraints. Making participatory ideals a norm requires dismantling systemic barriers through grand interventions, such as AI regulation. Without this, we face an ethical glass ceiling—able to make incremental progress yet blocked from transformative change.}

\edit{We find that, on one hand, each archetype has its own role to play in realizing a participatory future—even recognizing this as their  ``duty.''  On the other hand, navigating these roles is overwhelmingly frustrating, laborious, and littered with systemic barriers. The issue is not just that participatory ideals are lacking, but that they are structurally unachievable in the status quo.}

\section{Limitations}
We used non-random stratified sampling to recruit twelve participants across our roles of interest (researchers, developers, activists, and data workers). This sample size is limited and not comprehensive. For example, we had no participants who identified as genderqueer or gender non-conforming, despite recent efforts to create gender-diverse participatory AI environments~\cite{Queerinai2023-xy}. As with most interpretivist work, our methods are designed to create findings that prioritize transferability rather than generalizability~\cite{Soden2024-cf}. We find common ideals across many of our interviewees despite their different roles, identities, and experiences, indicating that our results transfer to various contexts. \edit{Following design traditions, we treat archetypes like personas—composite representations that guide decision-making rather than exhaustive classifications~\cite{Nielsen2019-ve}. AI encompasses far more than three archetypes, and articulating all possibilities would require a series of papers.} Future work could explore the concept of AI Participation through more scalable techniques such as large-scale surveying, discourse analysis of online conversations, or quantitative methods of participant impact on datasets.

\section{Conclusion}
In sum, we empirically investigate the experience of being a secondary stakeholder in the current age of AI. We start from the position that participatory AI efforts have laid important groundwork for building more ethical AI systems. However, there is a current gap between applying PAI methods to primary stakeholders and applying PAI ideals to secondary stakeholders. We seek to close this gap by first establishing three general ideals of meaningful participation: informedness, consent, and agency. We use these ideals as a lens to analyze 12 semi-structured interviews with secondary AI stakeholders. We find that there are important archetypes to consider, each with its own orientation toward and struggle with meaningful participation. For example, AI practitioners are attempting to broker more meaningful AI relationships but struggle to do so at the scale required to inform data contributors. We discuss how practitioners can use our archetypes as a roadmap for furthering both the empirical and theoretical study of AI participation.

\section{Statements}
\textbf{Ethical Considerations Statement. }This paper is human-subjects work and was approved by the University IRB \#00020382. In the spirit of meaningful participation, we heavily considered how to broker informedness, consent, and agency among all of our interviewees. In addition to written consent forms before the interview, we verbally explained the project in colloquial terms and reaffirmed consent. At both the beginning and end of each interview we reminded interviewees that they could withdraw their consent at any time or redact any information. Furthermore, we engaged with member checking strategies including communicating with participants after their interviews to confirm our own interpretations.

\textbf{Positionality Statement. }All of the authors of this work identify as well-intentioned practitioners. We all have experience with both human-subjects research and using data from public sites, such as social media. Therefore, it is important to note that the researchers of this work have reaped the benefits of the broad consent procedures and anti-agentic practices we critique in this paper. In keeping with member-research techniques, we believe our own struggles with realizing participatory ideals allowed us to appreciate our participants' experiences rather than overshadow them.

\textbf{Adverse Impact Statement.} Misinterpretations of our work could lead to a dilution of meaningful participation in AI systems. We are not suggesting that ameliorating the frustrations of data contributors, activists, and practitioners will solve all structural problems with AI systems. Therefore, adversarial actors could use our paper as a ``moral magic'' rather than a provocation for richer stakeholder consideration.

\begin{acks}
We are grateful to all of our participants for sharing their stories. Thank you to Nick Vincent and our reviewers for their feedback. This work was supported by the National Science Foundation (NSF) under Award \#2332841.
\end{acks}

\bibliographystyle{ACM-Reference-Format}
\bibliography{main}
\onecolumn
\clearpage
\appendix
\section{Participant Demographics}

\begin{table}[h!]
\begin{tabular}{|rll|rll}
\hline
\multicolumn{1}{l|}{\textbf{Demographic Variables}}             & \multicolumn{1}{l|}{\textbf{N}} & \textbf{Percentage}      & \multicolumn{1}{l|}{\textbf{Demographic Variables}}              & \multicolumn{1}{l|}{\textbf{N}} & \textbf{Percentage}      \\ \hline
\rowcolor[HTML]{A8D9E8} 
\multicolumn{1}{l}{\cellcolor[HTML]{A8D9E8}Age}                 &                                 &                          & \multicolumn{1}{l}{\cellcolor[HTML]{A8D9E8}Employment Status}    &                                 &                          \\ \hline
\multicolumn{1}{r|}{18-20}                                      & \multicolumn{1}{l|}{0}          & 0\%                      & \multicolumn{1}{r|}{Employed full-time}                          & \multicolumn{1}{l|}{5}          & 41\%                     \\
\multicolumn{1}{r|}{21-29}                                      & \multicolumn{1}{l|}{5}          & 41\%                     & \multicolumn{1}{r|}{Employed part-time}                          & \multicolumn{1}{l|}{1}          & 8\%                      \\
\multicolumn{1}{r|}{30-39}                                      & \multicolumn{1}{l|}{4}          & 33\%                     & \multicolumn{1}{r|}{Self-employed}                               & \multicolumn{1}{l|}{2}          & 17\%                     \\
\multicolumn{1}{r|}{40-49}                                      & \multicolumn{1}{l|}{3}          & 25\%                     & \multicolumn{1}{r|}{Student}                                     & \multicolumn{1}{l|}{3}          & 25\%                     \\ \cline{1-3}
\multicolumn{1}{l}{\cellcolor[HTML]{A8D9E8}Gender}              & \cellcolor[HTML]{A8D9E8}        & \cellcolor[HTML]{A8D9E8} & \multicolumn{1}{r|}{Unemployed (currently looking for work)}     & \multicolumn{1}{l|}{1}          & 8\%                      \\ \hline
\multicolumn{1}{r|}{Female}                                     & \multicolumn{1}{l|}{6}          & 50\%                     & \multicolumn{1}{l}{\cellcolor[HTML]{A8D9E8}Income}               & \cellcolor[HTML]{A8D9E8}        & \cellcolor[HTML]{A8D9E8} \\ \cline{4-6} 
\multicolumn{1}{r|}{Male}                                       & \multicolumn{1}{l|}{6}          & 50\%                     & \multicolumn{1}{r|}{\$0}                                         & \multicolumn{1}{l|}{1}          & 8\%                      \\ \cline{1-3}
\multicolumn{1}{l}{\cellcolor[HTML]{A8D9E8}Sexuality}           & \cellcolor[HTML]{A8D9E8}        & \cellcolor[HTML]{A8D9E8} & \multicolumn{1}{r|}{$1 - $9,999}                                 & \multicolumn{1}{l|}{2}          & 17\%                     \\ \cline{1-3}
\multicolumn{1}{r|}{Heterosexual}                               & \multicolumn{1}{l|}{10}         & 83\%                     & \multicolumn{1}{r|}{$10,000 - $24,999}                           & \multicolumn{1}{l|}{1}          & 8\%                      \\
\multicolumn{1}{r|}{Prefer not to say}                          & \multicolumn{1}{l|}{2}          & 17\%                     & \multicolumn{1}{r|}{$25,000 - $49,999}                           & \multicolumn{1}{l|}{3}          & 25\%                     \\ \cline{1-3}
\multicolumn{1}{l}{\cellcolor[HTML]{A8D9E8}Relationship Status} & \cellcolor[HTML]{A8D9E8}        & \cellcolor[HTML]{A8D9E8} & \multicolumn{1}{r|}{$50,000 - $74,999}                           & \multicolumn{1}{l|}{1}          & 8\%                      \\ \cline{1-3}
\multicolumn{1}{r|}{Single (never married)}                     & \multicolumn{1}{l|}{5}          & 41\%                     & \multicolumn{1}{r|}{$75,000 - $99,999}                           & \multicolumn{1}{l|}{3}          & 25\%                     \\
\multicolumn{1}{r|}{Married or Domestic Partnership}            & \multicolumn{1}{l|}{4}          & 33\%                     & \multicolumn{1}{r|}{\$150,000 and greater}                       & \multicolumn{1}{l|}{1}          & 8\%                      \\ \cline{4-6} 
\multicolumn{1}{r|}{In a relationship}                          & \multicolumn{1}{l|}{3}          & 25\%                     & \multicolumn{1}{l}{} &         &  \\ \hline
\multicolumn{1}{l}{\cellcolor[HTML]{A8D9E8}Ethnicity}           & \cellcolor[HTML]{A8D9E8}        & \cellcolor[HTML]{A8D9E8}  & \multicolumn{1}{r|}{ }                                  & \multicolumn{1}{l|}{}          &                       \\ 
\multicolumn{1}{r|}{African}                                    & \multicolumn{1}{l|}{2}          & 17\%                      & \multicolumn{1}{r|}{ }                                   & \multicolumn{1}{l|}{}          &                         \\
\multicolumn{1}{r|}{Black}                                      & \multicolumn{1}{l|}{2}          & 17\%                     & \multicolumn{1}{r|}{ }                                    & \multicolumn{1}{l|}{}          &                      \\
\multicolumn{1}{r|}{Hispanic}                                   & \multicolumn{1}{l|}{1}          & 8\%                      & \multicolumn{1}{r|}{ }                                & \multicolumn{1}{l|}{ }          &                       \\
\multicolumn{1}{r|}{South Asian}                                & \multicolumn{1}{l|}{2}          & 17\%                     & \multicolumn{1}{l|}{}                                            & \multicolumn{1}{l|}{}           &                          \\
\multicolumn{1}{r|}{Taiwanese}                                  & \multicolumn{1}{l|}{1}          & 8\%                      & \multicolumn{1}{l|}{}                                            & \multicolumn{1}{l|}{}           &                          \\
\multicolumn{1}{r|}{Indigenous}                                 & \multicolumn{1}{l|}{1}          & 8\%                      & \multicolumn{1}{l|}{}                                            & \multicolumn{1}{l|}{}           &                          \\
\multicolumn{1}{r|}{White}                                      & \multicolumn{1}{l|}{3}          & 25\%                     & \multicolumn{1}{l|}{}                                            & \multicolumn{1}{l|}{}           &                         
\end{tabular}
\caption{Aggregate participant demographics.}
\label{tab:demographics}
\end{table}

\begin{table*}[h!]
  \rowcolors{2}{gray!10}{white} 
  \begin{tabular}{ll}
    \toprule
    \textbf{Pseudonym} & \textbf{Role Description}  \\
    \midrule
    Maddie & Non-profit Industry AI Researcher \\ 
    Sean & Participatory AI Researcher in Agricultural Contexts \\  
    Tyler & Participatory AI Researcher Studying Online Communities \\  
    Scott & AI Developer Focused on Tools for Software Engineers\\  
    Falen & Expert Data Annotator for Commercial Mental Health AI \\  
    Sabrina & Social Computing Researcher and Volunteer Moderator \\ 
    Fiona & Former Commercial Content Moderator and Current Content Moderators Activist  \\
    Katie & Former mTurk Data Worker and Current Data Rights Activist\\
    Kayla & AI Ethics Researcher and Activist \\
    Charles & AI Developer for Company\\ 
    Anthony & AI Developer for Company \\  
    Spencer & AI Researcher Bulding Public Sector Tools  \\  
    
    \bottomrule
  \end{tabular}
  \caption{Interview participants' pseudonym and their role description. All names in this section are pseudonyms from the top 100 US baby names and, therefore, do not indicate the culture or ethnic background of the participants. Company and institution names have been redacted unless the participant explicitly requested otherwise.} 
  \label{tab:psuedonyms}
\end{table*}
\end{document}